\documentclass[12pt]{article}
\usepackage[OT1]{fontenc}
\usepackage[authoryear]{natbib}
\usepackage[colorlinks,citecolor=blue,urlcolor=blue]{hyperref}
\usepackage{amsthm,amsmath}
\usepackage{times,color,epsfig,subfig,layout,lscape,latexsym,amsfonts,amssymb,graphicx,rotating,setspace}
\usepackage{soul,xcolor}
\usepackage{xr}
\UseRawInputEncoding  
\newtheorem{theorem}{Theorem}  

\newcommand{\R}{\ensuremath{\mathbb{R}}}

\newcommand{\uca}{\ensuremath{c_{\mu}}}
\newcommand{\ucb}{\ensuremath{c_{\sigma^{2}}}}

\renewcommand{\vec}[1]{\boldsymbol{#1}}

\renewcommand{\baselinestretch}{1.2}

\def\tit{Hypothesis Testing for Two Sample Comparison of Network Data}

\def\abst{Network data is a major object data type that has been widely collected or derived from common sources such as brain imaging. Such data contains numeric, topological, and geometrical information, and may be necessarily considered in certain non-Euclidean space for appropriate statistical analysis. The development of statistical methodologies for network data is challenging and currently at its infancy; for instance, the non-Euclidean counterpart of basic two-sample tests for network data is scarce in literature. In this study, a novel framework is presented for two independent sample comparison of networks. Specifically, an approximation distance metric to quotient Euclidean distance is proposed, and then combined with network spectral distance to quantify the local and global dissimilarity of networks simultaneously. A permutational non-Euclidean analysis of variance is adapted to the proposed distance metric for the comparison of two independent groups of networks. Comprehensive simulation studies and real applications are conducted to demonstrate the superior performance of our method over other alternatives. The asymptotic properties of the proposed test are investigated and its high-dimensional extension is discussed as well.}

\def\keyword{Object Data; Brain Connectivity Network; Hypothesis Testing in Non-Euclidean Space; Quotient Ipsen-Mikhailov Distance; Permutational Non-Euclidean Two-sample Test. }

\def\authorlist{Han Feng, Xing Qiu, Hongyu Miao}

\begin{document}

\setlength{\textheight}{8.8in}
\setlength{\textwidth}{6in}
\setlength{\topmargin}{-36pt}
\setlength{\oddsidemargin}{0pt}
\setlength{\evensidemargin}{0pt}
\tolerance=500

\def\omitfirst
{
\begin{center}
{\large \bf \tit} \footnote{Hongyu Miao (E-mail: {\it
hongyu.miao@uth.tmc.edu}) is Associate Professor,
Han Feng (E-mail: {\it han.feng@uth.tmc.edu}) is PhD Student,
Department of Biostatistics and Data Science, University of Texas Health Science Center at Houston.
Xing Qiu (E-mail: {\it xing\_qiu@urmc.rochester.edu}) is Professor,
Department of Biostatistics and Computational Biology, University of
Rochester.}

\vspace*{0.3cm}
{\authorlist}
\end{center}

\makeatletter

\thispagestyle{empty}

\begin{abstract}
\abst
\end{abstract}

\noindent{\bf Key Words}: \keyword \\

\noindent{\bf Short Title}: Two Sample Test for Network Data

\newpage\thispagestyle{empty}

\begin{center}
{\large \bf \tit}
\end{center}

\vspace*{0.4cm}

\begin{abstract}
\abst
\end{abstract}
\noindent{\bf Key Words}: \keyword
\clearpage\pagebreak\newpage
\pagenumbering{arabic}
}

\omitfirst

\setcounter{page}{1}

\pagenumbering{arabic}

\begin{doublespace}

\section{Introduction} \label{IntroSect}


Object data refers to a broad range of complex data that contains internal structures and has drawn increasing attention of the statistical and data science communities \citep{strogatz2001exploring, hoff2002latent, newman2003structure, wang2007object, marron2014overview, lu2014object, menafoglio2017statistical}. 
Network is a major type of object data, which consists of nodes, edges and associated attributes, and contains rich numeric, topological, and geometrical information. Typical examples of such data include World Wide Web, social networks, traffic networks, gene regulatory networks, and neural networks \citep{park2003hyperlink, borgatti2009network, paliwal2009neural, marbach2010revealing, tian2016analysis}. 
Network data can be directly observed or derived from other data types; for instance, brain connectivity networks are inferred from functional magnetic resonance imaging (fMRI) data, and have been frequently investigated for novel scientific insights \citep[e.g.,][]{greicius2003functional, supekar2008network, bullmore2012economy, Jiang2020connect}. 
In this study, we use brain connectivity networks as the primary application example; however, the proposed methodology is generally applicable to other network data.



Most existing statistical methodologies for network analysis focus on inferring internal network structures from static or time-course data \citep[e.g.,][]{gui2017} or quantitatively characterizing various graph properties \citep{costa2007characterization, kolaczyk2014statistical}. Due to a variety of technical challenges in parameterization and analysis, networks have been primarily considered as elements in high-dimensional Euclidean spaces for simplicity \citep{kolaczyk2014statistical, ginestet2017hypothesis, gao2019geometric, biswas2020covid}. 
However, previous work has shown that even for simple network data like trees, the data space could be strongly non-Euclidean \citep{wang2007object, nye2011principal} such that conventional statistical methods formulated in Euclidean space are sub-optimal or even inapplicable.
For example, in recent years, a few two-sample hypothesis testing procedures for network data have been proposed based on Euclidean distance metrics. The work of \cite{ginestet2014statistical} was developed based on the global efficiency measure proposed by \cite{latora2001efficient} for networks with the small-world property, which is an assumption that may not hold in many situations. The work of \cite{ginestet2017hypothesis} presented an alternative method for network comparisons based on the Hotelling's $T^2$ test idea. However, this approach oversimplifies the problem and the optimal performance of the proposed test can only be achieved under certain restrictive assumptions (e.g., equal variance-covariance matrix assumption on the two groups of edge weights). \cite{Chen2017} proposed an edge-count test based on the pairwise distances calculated through minimum spanning tree and k-nearest neighbor. As a nonparametric testing approach, this method is flexible and applicable to network data and does not depend on the assumptions that other previous work requires. However, Chen's method is still Euclidean in spirit; and it only captures the local structural difference of networks and overlooks global network properties, which may lead to the loss of efficiency for two-sample comparisons.

While not specifically for network data, \cite{anderson2013permanova} proposed the permutational multivariate analysis of variance (PERMANOVA) method that allows the incorporation of arbitrary non-Euclidean distances. However, the performance of PERMANOVA deteriorates with the increase in data heteroscedasticity \citep{warton2012distance}. Based on PERMANOVA, \cite{Alekseyenko2016} proposed a multivariate Welch t-test that utilized pairwise distances to conduct between-group comparisons via (pseudo) F-statistics. Similar to PERMANOVA, the multivariate Welch t-test can also accommodate non-Euclidean distances. Note that the non-Euclidean distance metrics mentioned in \cite{anderson2013permanova} and \cite{Alekseyenko2016} are very general and not tailored to any specific object data. In deed, it is a challenging task to define efficient and accurate distance metrics for object data in non-Euclidean space; for instance, only for tree-structure data, \cite{Feragen2013} proposed a rigorous but computationally-expensive non-Euclidean distance metric, called quotient Euclidean distance (QED).




Motivated by the pioneer studies above,  we propose a novel non-Euclidean and nonparametric framework for comparing two independent samples of networks with high efficiency.  Particularly, differences in local network structure are quantified by an innovative pseudo distance metric in quotient Euclidean space, and the global network difference is quantified by the spectral distance. The combination of the local and global network distances leads to an efficient non-Euclidean metric for measuring network dissimilarity. Furthermore, this distance metric enables us to adapt the permutational non-Euclidean ANOVA $F$-test for comparing two groups of networks. The validity and efficiency of the proposed two-sample test are verified through both theoretical justifications and extensive simulation studies. In particular, we derive several universal asymptotic properties of the $F$-statistic under the null and alternative hypotheses for a large class of distance metrics. Interestingly, we also show that while the first and second order moments of the distance matrix enjoy certain universal properties, the \emph{shape} of the distribution do not. It strongly depends on the geometric and distributional properties of the networks and the choice of the distance function. Therefore, it is theoretically impossible to pursue one universal limit distribution and asymptotic relative efficiency (ARE) for the distance-based $F$-statistic for all distance functions in Euclidean or non-Euclidean spaces. Finally, we also discuss possible extensions of the proposed method to high dimensional cases.

To the best of our knowledge, this study is among the very few pioneer efforts that rigorously address the two independent sample test problem for network data in a non-Euclidean (and often high-dimensional) space. This article is organized as follows. The details of the proposed methodology are described in Section~\ref{ModelSect}. Asymptotic properties of our method are investigated in Section~\ref{TheorySect}. We conduct multiple simulation experiments to evaluate the performances of the proposed method and other alternatives in Section~\ref{SimuSect}. The use of the proposed test in practice is illustrated by a real data example in Section~\ref{AppSect}. Finally, we discuss the advantages and limitations of the proposed test, as well as its extension to high dimensional cases in Section~\ref{DiscConcSect}.

\section{Method}\label{ModelSect}

\subsection{Parameterization and Distance Metric}\label{DistDefSect}
A typical parameterization of network topological structure is $\mathcal{G}=\left\{ \mathcal{V}, \mathcal{E} \right\}$, where $\mathcal{V}$ denotes the set of nodes and $\mathcal{E}$ denotes the set of edges. The edges in $\mathcal{E}$ can be directed or undirected, depending on specific problems. In this study, we only consider simple networks; that is, no self-loops are allowed and there exists at most one edge between any pair of nodes. In addition to topological structure, a network $\mathcal{G}$ may also have attributes associated with its nodes and/or edges. Let $\mathcal{X}_{\mathcal{V}}$ and $\mathcal{X}_{\mathcal{E}}$ denote the attribute sets associated with $\mathcal{V}$ and $\mathcal{E}$, respectively, then a more comprehensive parameterization of network data is $\mathcal{G}=\left\{ \mathcal{V}, \mathcal{E}, \mathcal{X}_{\mathcal{V}}, \mathcal{X}_{\mathcal{E}} \right\}$. When node attributes do not exist or are not of interest, this parameterization can be further simplified to the adjacency matrix representation. If the elements in $\mathcal{X}_{\mathcal{E}}$ are $q$-dimensional vectors ($q>1$), the adjacency matrix is actually a $v \times v \times q$ dimensional tensor, where $v=\left| \mathcal{V} \right|$ is the cardinality of $\mathcal{V}$. For simplicity, we assume $q=1$ in this study so the adjacency matrix is a $v \times v$ matrix, with zero elements denoting nonexistence of edges and non-zero elements being the attribute values of the corresponding edges, denoted by $\mathbf{W} \in \mathcal{R}^{v \times v}$.

\begin{figure}
{\includegraphics[width=1\textwidth]{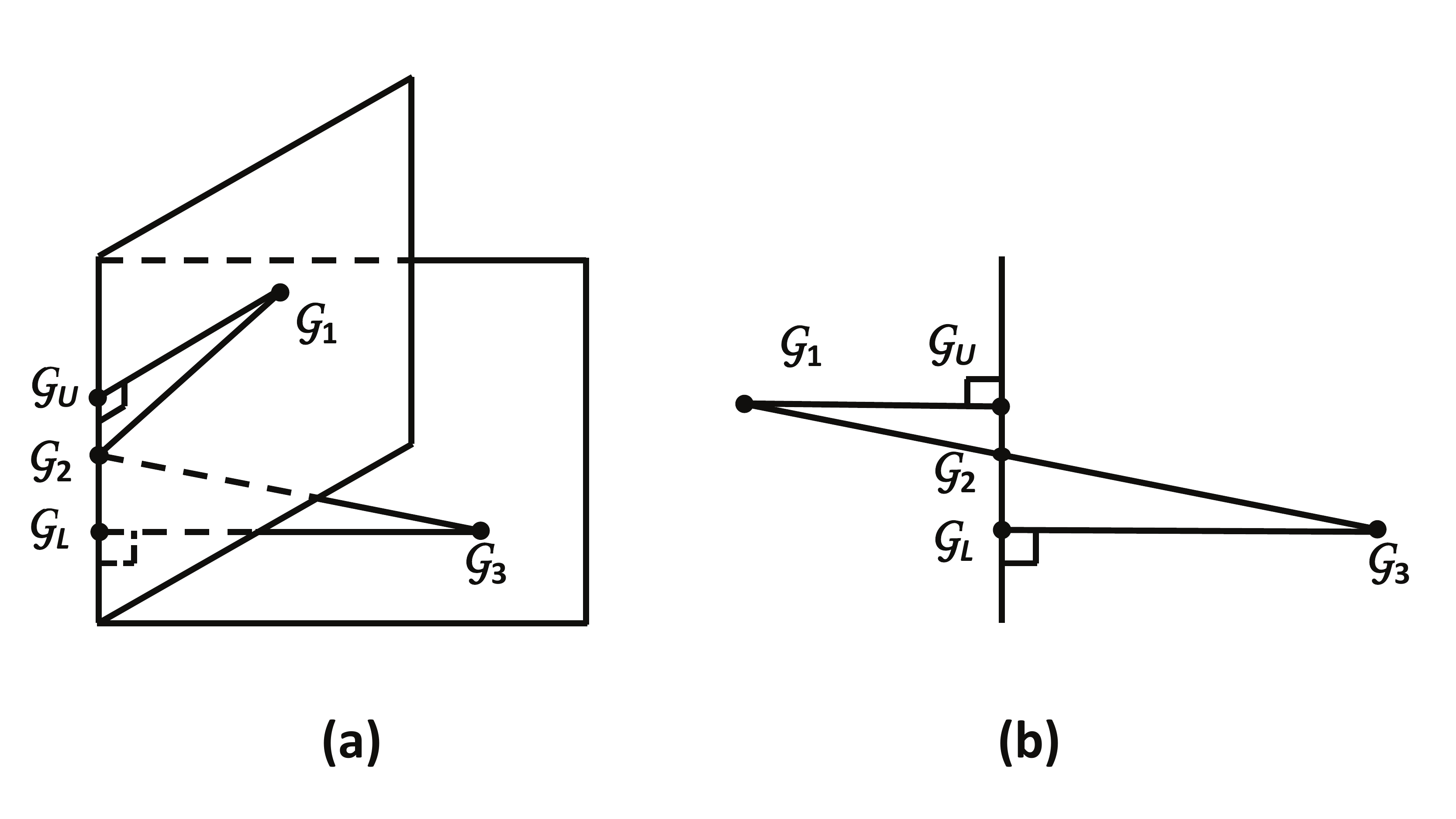}}
\caption{Illustration of the quotient space geometry, where $\mathcal{G}_1 \sim \mathcal{G}_2$ and $\mathcal{G}_2 \sim \mathcal{G}_3$ but $\mathcal{G}_1 \nsim \mathcal{G}_3$. (a) A feasible path from $\mathcal{G}_1$ to $\mathcal{G}_3$ is via $\mathcal{G}_2$ because $\mathcal{G}_2$ is on the boundary of the two Euclidean spaces. (b) Transition pathes from $\mathcal{G}_1$ to $\mathcal{G}_3$ after unfolding the two Euclidean hyperplanes.}
\label{figQuotientSpace}
\end{figure}

Given the topological structure of $\mathcal{G}$, it is more appropriate to analyze $\mathbf{W}$ in a quotient space \citep{Bridson1999}. First, consider two network data points $\mathcal{G}_1$ and $\mathcal{G}_2$ that have the same set of nodes and edges but differ from each other in edge attributes. We call that $\mathcal{G}_1$ is coplanar to $\mathcal{G}_2$, denoted by $\mathcal{G}_1 \sim \mathcal{G}_2$ with $\sim$ representing a coplane relation \citep{Feragen2013}. Considering the simplest case that $\mathcal{G}_1$ and $\mathcal{G}_2$ differ from each other in only one edge attribute, the distance between $\mathcal{G}_1$ and $\mathcal{G}_2$ is just the Euclidean distance $\left| \mathcal{X}_{\mathcal{E}_{ij}} (\mathcal{G}_1) - \mathcal{X}_{\mathcal{E}_{ij}} (\mathcal{G}_2) \right| = \left| \mathbf{W}_1(i,j) - \mathbf{W}_2(i,j) \right|$, where $\mathcal{X}_{\mathcal{E}_{ij}} (\mathcal{G})$ denotes the attribute of edge $\mathcal{E}_{ij}$ from node $j$ to node $i$ in network $\mathcal{G}$, and $\mathbf{W}(i,j)$ denotes the element at the $i$-th row and the $j$-th column. Thus, coplanar $\mathcal{G}$s are in the same Euclidean space, and the distance metric is the regular Euclidean distance like the Frobenius norm $d(\mathbf{W}_1,\mathbf{W}_2) = \sqrt{\Sigma_{i,j} \left( \mathbf{W}_1(i,j) - \mathbf{W}_2(i,j) \right)^2}$. Second, consider three data points $\mathcal{G}_1$, $\mathcal{G}_2$ and $\mathcal{G}_3$, where $\mathcal{G}_1 \sim \mathcal{G}_2$ and $\mathcal{G}_2 \sim \mathcal{G}_3$ but $\mathcal{G}_1 \nsim \mathcal{G}_3$. This can happen if an edge vanishes when its certain attribute(s) becomes zero. For example, if only consider network topology for now, $\mathcal{G}_1$, $\mathcal{G}_2$ and $\mathcal{G}_3$ share the same set of nodes, and $\mathcal{G}_1$ or $\mathcal{G}_3$ can be transformed to $\mathcal{G}_2$ by removing only one edge. Thus, $\mathcal{G}_2$ is on the boundary of the Euclidean space where the coplanar classes of $\mathcal{G}_1$ or $\mathcal{G}_3$ lie in. This conclusion can also be drawn if we look at the weighted adjacency matrices: one of the non-zero elements $\mathbf{W}_1(i,j)$ in $\mathbf{W}_1$ and one of the non-zero elements $\mathbf{W}_3(k,l)$ in $\mathbf{W}_3$ become zero in $\mathbf{W}_2$ ($i \neq k$ and/or $j \neq l$). Therefore, for a set of network data points that share the same set of nodes, the corresponding quotient space consists of a number of piecewise Euclidean spaces stitched together, and it is non-differentiable on the boundaries of any two piecewise Euclidean spaces. Such a geometry of the quotient space suggests that for two network data points that are not coplanar, a feasible path between them must be along (multiple) pieces of Euclidean spaces and cross their boundaries; for instance, one needs to transform $\mathcal{G}_1$ to $\mathcal{G}_3$ via $\mathcal{G}_2$, see Fig. \ref{figQuotientSpace}(a) for illustration. Hence, the geodesic is the length of the shortest feasible path(s) between two data points, called the quotient Euclidean distance (QED) \citep{Feragen2013}.

The definition of QED is consistent with the long standing definition of graph edit distance \citep{Sanfeliu1983}; however, the calculation of QED is very challenging because there may exist numerous paths between two network data points such that it becomes impractical to enumerate all of them to find the shortest one \citep{LuMiao2016}. Therefore, an efficient approximation to QED is necessary. As shown in Fig. \ref{figQuotientSpace}(b), after unfolding the two neighboring Euclidean spaces, the QED between $\mathcal{G}_1$ and $\mathcal{G}_3$ is
\begin{equation}\label{eqnQED13}
  QED(\mathcal{G}_1, \mathcal{G}_3) = d(\mathcal{G}_1, \mathcal{G}_2) + d(\mathcal{G}_2, \mathcal{G}_3),
\end{equation}
where $d(\cdot)$ denotes a feasible distance metric. The weighted adjacency matrix has a natural chart $\varphi : \mathbf{W}_{v \times v} \rightarrow \vec{w}_{v^2 \times 1}$, where $\vec{w}_{v^2 \times 1}$ denotes a $v^2$-dimensional vector obtained by stacking the columns of $\mathbf{W}$ one above another. Then the QED approximation $d_a$ considered here is
\begin{equation}\label{eqnQEDapprox}
  d_a (\mathcal{G}_1, \mathcal{G}_3) = \left|\vec{w}_1 - \vec{w}_3\right|,
\end{equation}
where $\left| \cdot \right|$ denotes a $L^1$ norm, and $\vec{w}_1$ and $\vec{w}_3$ are generated from $\mathcal{G}_1$ and $\mathcal{G}_3$, respectively. Note that this approximation corresponds to multiple feasible pathes, and it has been previously shown that the approximation error of $d_a (\mathcal{G}_1, \mathcal{G}_3)$ is bounded by the following distance \citep{LuMiao2016}
\begin{equation}\label{eqnQEDapproxbound}
  d_a (\mathcal{G}_1, \mathcal{G}_3) = d(\mathcal{G}_1, \mathcal{G}_U) + d(\mathcal{G}_U, \mathcal{G}_2) + d(\mathcal{G}_2, \mathcal{G}_L) + d(\mathcal{G}_L, \mathcal{G}_3),
\end{equation}
where $\mathcal{G}_U$ and $\mathcal{G}_L$ denote the projections of $\mathcal{G}_1$ and $\mathcal{G}_3$ onto the boundary, respectively (see Fig. \ref{figQuotientSpace}).

An alternative interpretation of the QED approximation is that it measures the total cost associated with the presence and absence of matching edges, given that two networks sharing the same nodes. Considering edges as independent entities, the QED approximation is actually a local measure of dissimilarity between networks as it ignores the overall network structure characteristics. Particularly, the Ipsen-Mikhailov distance \citep{Ipsen2003} has been introduced as a global measure although it cannot distinguish isospectral or isomorphic networks. Specifically, an $n$-nodes network can be treated as a system of $n$ balls connected by elastic springs, and its vibration frequencies $\psi_i$ are the square roots of the Laplacian matrix eigenvalues ($\epsilon_1=0$, $\psi_i = \sqrt{\epsilon}_i$, $i=2,...,n$). Since the spectral density $\rho (\psi)$ of an arbitrary graph does not necessarily follow the semi-circle law \citep{Farkas2001}, it is more general to consider it in forms of the sum of Lorentz distributions \citep{Ipsen2004}
\begin{equation}\label{eqnSpectralDensity}
  \rho (\psi, \gamma) = K \sum_{i=2}^{n} \frac{\gamma}{\left( \psi - \psi_i \right)^2 + \gamma^2},
\end{equation}
where $K$ is the normalization constant and $\gamma$ is the common width. Let $\gamma^*$ denote the unique solution \citep{Jurman2015} to
\begin{equation}\label{eqnEmptyFullSpectral}
\sqrt{\int_{0}^{\infty} \left[ \rho_{e}(\psi, \gamma)-\rho_{f}(\psi, \gamma) \right]^2 d \psi} = 1,
\end{equation}
where $\rho_{e}$ is the spectral density of an empty network with $n$ nodes and zero edges and $\rho_{f}$ is the spectral density of an undirected simple full network with $n$ nodes and $n^2$ edges. Then the normalized Ipsen-Mikhailov distance is defined as
\begin{equation}\label{eqnIMdist}
  IM(\mathcal{G}_1, \mathcal{G}_3) = \sqrt{\int_{0}^{\infty} \left[ \rho_1(\psi, \gamma^*)-\rho_3(\psi, \gamma^*) \right]^2 d \psi},
\end{equation}
which is bounded between 0 and 1. \cite{Jurman2015} suggested to combine the local Humming edit distance \citep{Morris2008} with the global Ipsen-Mikhailov distance to form the glocal Hamming-Ipsen-Mikhailov (HIM) distance for network comparison and classification. Since the Humming edit distance is just a convenient simplification of the proposed QED approximation, the spectral distance is combined with the QED approximation in this study. The final distance metric, called Quotient Ipsen-Mikhailov (QIM) distance, becomes
\begin{equation}\label{eqnQIMmetric}
  d_q(\mathcal{G}_1, \mathcal{G}_3) = \left|\vec{w}_1 - \vec{w}_3\right| \cdot \left[ 1 + \kappa \cdot \sqrt{\int_{0}^{\infty} [\rho_1(\psi, \gamma^*)-\rho_3(\psi, \gamma^*)]^2 d \psi} \right],
\end{equation}
where $\kappa$ is a constant that represents the weight of global dissimilarity. The rationale of Eq. (\ref{eqnQIMmetric}) is that a high local similarity of two networks (i.e., a small QED approximation) suggests a higher global similarity (i.e., a small Ipsen-Mikhailov distance), but not vice versa.

\subsection{Two Independent Sample Test}\label{TestDefSect}
Once a distance metric is defined, test statistics can be developed for two independent sample comparison, which is the focus of this section. Let $\mathcal{A}=\left\{ \mathcal{G}_i^{\mathcal{A}} \right\}_{i=1}^{n_{A}}$ and  $\mathcal{B}=\left\{ \mathcal{G}_j^{\mathcal{B}} \right\}_{j=1}^{n_{B}}$ denote two samples from network space, and assume $\mathcal{G}_1^{\mathcal{A}}, \dots, \mathcal{G}_{n_{A}}^{\mathcal{A}}$ are i.i.d. samples from a probability distribution $P_{A}$ and $\mathcal{G}_1^{\mathcal{B}}, \dots, \mathcal{G}_{n_{B}}^{\mathcal{B}}$ are i.i.d. from $P_{B}$. Let $\Omega$ denote the space of a family of distribution pairs $(P_{A},P_{B})$ that adopts a nonparametric model. Let $n=n_{A}+n_{B}$, and let $\mathcal{Z}$ denote the set of all data points
$$\mathcal{Z}=\left( \mathcal{G}_1^{\mathcal{Z}}, \dots, \mathcal{G}_n^{\mathcal{Z}} \right) = \left( \mathcal{G}_1^{\mathcal{A}}, \dots, \mathcal{G}_{n_{A}}^{\mathcal{A}},\mathcal{G}_1^{\mathcal{B}}, \dots, \mathcal{G}_{n_{B}}^{\mathcal{B}}\right).$$
It is straightforward to devise a test statistic like $d_q(\hat{\mu}_{\mathcal{A}}, \hat{\mu}_{\mathcal{B}})$, where $\hat{\mu}_{\mathcal{A}}$ and $\hat{\mu}_{\mathcal{B}}$ are the Fr\'{e}chet means \citep{Frechet1948} of $\mathcal{A}$ and $\mathcal{B}$, respectively.

However, there are several hurdles to overcome here. For instance, the calculation of Fr\'{e}chet mean of networks is challenging and only a few algorithms exist for specific types of networks (i.e., trees) \citep{Sturm2003, Bacak2014, Miller2015}. In addition, for hypothesis testing of equal variance between $P_{A}$ and $P_{B}$, the calculation of network variance is necessary but remains an open problem \citep{Saerens2004, Shahid2016}.

Instead, the test statistic can be derived by comparing the within-sample $d_q$ with the between-sample $d_q$. Note that this two sample test is built upon $d_q$, which is a non-Euclidean distance; therefore, permutational non-Euclidean analysis of variance \citep{Alekseyenko2016} technique can be considered and the test statistic is thus given as follows
\begin{equation}\label{eqnTest}
  F = \frac{\frac{1}{n} \cdot \sum\limits_{\substack{i,j=1 \\ i<j}}^{n} d_q^2(\mathcal{G}_i^{\mathcal{Z}}, \mathcal{G}_j^{\mathcal{Z}}) - \frac{1}{n_{A}} \cdot \sum\limits_{\substack{i,j=1 \\ i<j}}^{n_{A}} d_q^2(\mathcal{G}_i^{\mathcal{Z}}, \mathcal{G}_j^{\mathcal{Z}}) - \frac{1}{n_{B}} \cdot \sum\limits_{\substack{i,j=n_{A}+1 \\ i<j}}^{n} d_q^2(\mathcal{G}_i^{\mathcal{Z}}, \mathcal{G}_j^{\mathcal{Z}})} {\frac{1}{n-2} \cdot \left( \frac{1}{n_{A}} \cdot \sum\limits_{\substack{i,j=1 \\ i<j}}^{n_{A}}d_q^2(\mathcal{G}_i^{\mathcal{Z}}, \mathcal{G}_j^{\mathcal{Z}})  + \frac{1}{n_{B}} \cdot \sum\limits_{\substack{i,j=n_{A}+1 \\ i<j}}^{n} d_q^2(\mathcal{G}_i^{\mathcal{Z}}, \mathcal{G}_j^{\mathcal{Z}}) \right)}.
\end{equation}
The exact distribution $F$ is difficult to specify, so a permutation test procedure is considered for the null hypothesis $\text{H}_0: P_{A}=P_{B}=P_{0}$ v.s. the alternative hypothesis $\text{H}_1: P_{A} \neq P_{B}$. Let $\pi_k = \left( \pi_k(1), \dots, \pi_k(n) \right)$, $k=1,\dots,K$, denote the $K$ permutations of $\left\{1,\dots,n\right\}$, and let $F_{k} = F\left( \mathcal{G}_{\pi_k(1)}^{\mathcal{Z}}, \dots, \mathcal{G}_{\pi_k(n)}^{\mathcal{Z}} \right)$ denote the corresponding test statistic. Comparing the $F_{k}$ with the original test score $F_{0}=F(\mathcal{G}_1^{\mathcal{Z}}, \dots, \mathcal{G}_n^{\mathcal{Z}})$, the $p$-value can be obtained as follows
\begin{equation}\label{eqnPvalue}
  p= \frac{c + \sum\limits_{k=1}^{K} \mathcal{I}\left\{ F_{k} \geq F_{0} \right\}}{K+c},
\end{equation}
where $\mathcal{I}\left\{\cdot\right\}$ is the indicator function and $c$ is a small constant that is added to avoid $p=0$ \citep{Knijnenburg2009}.

\section{Theoretical Properties of the Proposed Test}\label{TheorySect}

\subsection{Assumptions}
\label{sec:notations-assumptions}

For the clarity of derivations in this section, we denote the $n\times n$-dimensional squared pairwise distance matrix between all samples by $\mathcal{D}$, i.e., $\mathcal{D}_{i,j} := d^{2}(\mathcal{G}^{\mathcal{Z}}_{i}, \mathcal{G}^{\mathcal{Z}}_{j})$. When the sample size $n$ needs to be emphasized, e.g., in large sample derivations, $\mathcal{D}$ may be written as $\mathcal{D}_{n}$. Using this notation, the $F$-statistic defined in Equation~\eqref{eqnTest} becomes
\begin{equation}\label{eqnFTest}
  \begin{gathered}
    F(\mathcal{D}) = \frac{ \frac{1}{n_{A}+n_{B}} \cdot \sum\limits_{\substack{i,j=1 \\ i<j}}^{n_{A}+n_{B}} \mathcal{D}_{ij} - \frac{1}{n_{A}} \cdot \sum\limits_{\substack{i,j=1 \\ i<j}}^{n_{A}} \mathcal{D}_{ij} - \frac{1}{n_{B}} \cdot \sum\limits_{\substack{i,j=n_{A}+1 \\ i<j}}^{n_{A}+n_{B}} \mathcal{D}_{ij} } { \frac{1}{n-2} \cdot \left( \frac{1}{n_{A}} \cdot \sum\limits_{\substack{i,j=1 \\ i<j}}^{n_{A}} \mathcal{D}_{ij}  + \frac{1}{n_{B}} \cdot \sum\limits_{\substack{i,j=n_{A}+1 \\ i<j}}^{n_{A}+n_{B}} \mathcal{D}_{ij} \right)}.
  \end{gathered}
\end{equation}

Let $\mathrm{sym}(n)$ be the symmetric group that contains all permutations of $n$ symbols. For a given $\mathcal{D}$ and $\pi \in \mathrm{sym}(n)$, the permuted distance matrix, denoted by $\pi(\mathcal{D})$, is constructed by permuting both the row and columns of $\mathcal{D}$ by $\pi$. In other words, $[\pi(\mathcal{D})]_{i,j} := \mathcal{D}_{\pi(i),\pi(j)}$.

We require the following two additional assumptions in this section:
\begin{enumerate}
\item \textbf{Symmetry of within-group distance}: The marginal distribution of the squared distance between two networks in the same group, $\mathcal{D}_{i,j} := d^{2}(\mathcal{G}^{\mathcal{Z}}_{i}, \mathcal{G}^{\mathcal{Z}}_{j})$ is the same irrespective of whether both $i,j \in A$ or $i,j \in B$. In other words, switching the group of $\mathcal{G}^{\mathcal{Z}}_{i}$ and $\mathcal{G}^{\mathcal{Z}}_{j}$ from A to B or B to A does not change statistical properties of within-group distance $\mathcal{D}_{i,j}$.  \label{assumption:symmetry}
\item \textbf{Finite moments}: When viewed as a random matrix, $\mathcal{D}$ has finite first and second order moments. \label{assumption:finite-moments}
\end{enumerate}

\subsection{Type I Error}\label{Type1Sect}

Let
\begin{equation*}
  F^{(1)}(\mathcal{D}) \leqslant F^{(2)}(\mathcal{D}) \leqslant \dots \leqslant F^{(n!)}(\mathcal{D})
\end{equation*}
be the \emph{order statistics} of $\left\{F(\pi(\mathcal{D})_{n}): \pi\in \mathrm{sym}(n) \right\}$, the set of test statistics computed from all permuted samples.

For a given significance level $\alpha \in (0,1)$, we let $k = \lceil n!(1-\alpha) \rceil$ be the least integer greater than or equal to $n!(1-\alpha)$, and define
\begin{equation}
  \label{eqnMplut-M0}
  \begin{split}
    M^{+}(\mathcal{D}) &= \left|\left\{ 1\leqslant j \leqslant n! : F^{(j)}(\mathcal{D}) > F^{(k)}(\mathcal{D}) \right\} \right|, \\
    M^{0}(\mathcal{D}) &= \left|\left\{ 1\leqslant j \leqslant n! : F^{(j)}(\mathcal{D}) = F^{(k)}(\mathcal{D}) \right\} \right|,
  \end{split}
\end{equation}
as the number of the summary statistics computed from the permuted samples that are strictly greater than $F^{(k)}(\mathcal{D})$ and equal to $F^{(k)}(\mathcal{D})$, respectively.

With these notations, the permutation $F$-test for a distance matrix $\mathcal{D}$ can be defined formally as
\begin{equation}
  \label{eqnrandomized-test}
  \phi(\mathcal{D}) =
  \begin{cases}
    1, & \text{if $F(\mathcal{D}) > F^{(k)}(\mathcal{D})$}, \\
    a(\mathcal{D}), & \text{if $F(\mathcal{D}) = F^{(k)}(\mathcal{D})$}, \\
    0, & \text{if $F(\mathcal{D}) < F^{(k)}(\mathcal{D})$}.
  \end{cases} \qquad a(\mathcal{D}) := \frac{n!\cdot \alpha -M^{+}(\mathcal{D})}{M^{0}(\mathcal{D})}.
\end{equation}

\begin{theorem}\label{thmtype-I-error}
  The above defined permutation $F$-test for a distance matrix is \textbf{exact}. In other words, under $H_{0}$, the type I error of this test equals $\alpha$:
  \begin{equation}
    \label{eqnexactness}
    E_{P_{0}} \left( \phi(\mathcal{D}) \right)= \alpha.
  \end{equation}
  Here $P_{0}$ is the shared distribution of two samples under $H_{0}$.
\end{theorem}
\begin{proof}
  By construction, we know that the sample $\{\mathcal{G}^{\mathcal{Z}}_{1}, \dots, \mathcal{G}^{\mathcal{Z}}_{n}\}$ is exchangeable w.r.t. random permutation under $H_{0}$. Since distance $d(\cdot, \cdot)$ is a measurable function, $\mathcal{D} \stackrel{d}{=} \pi(\mathcal{D})$ for all $\pi \in \mathrm{sym}(n)$. Theorem~\ref{thmtype-I-error} is an immediate consequent of \cite[Theorem 15.2.1]{lehmann2006testing}.
\end{proof}

\subsection{Universal Asymptotic Properties of the Test Statistic}\label{UniPropSect}

Theorem~\ref{thmtype-I-error} guarantees that the proposed test has tight control of type I error irrespective of the underlying distribution of networks and the form of $d(\cdot, \cdot)$. Unfortunately, it is much more difficult to study the statistical power of the proposed permutation $F$-test, because there is no one universal asymptotic distribution of $F(\mathcal{D}_{n})$, when $n\to \infty$, that works for all distance function $d(\cdot, \cdot)$ and the underlying distribution of the networks. We will demonstrate this point in Section~\ref{CasePropSect}.

That being said, we are able to derive certain weaker universal asymptotic properties of $F(\mathcal{D})$ in terms of the first and second order moments. These conclusions are valid for all distance functions and generative models for networks that satisfy very mild conditions. As we all know, the first two moments of $F(\mathcal{D})$ provide not only a ``rough outline'' of its distribution, but also rigorous (albeit a little loose) upper bounds of its tail probability based on Chebyshev's inequality. These findings are summarized in Theorem~\ref{thm:main}.




Because the observed networks are $i.i.d.$ within each group and a distance function is always symmetric, the marginal distribution of $\mathcal{D}_{ij}$ can only take three forms, depending on whether the combination of the two input networks are of type AA, BB (both are within group), or AB (between-group). Based on Assumption~\ref{assumption:symmetry} in Section~\ref{sec:notations-assumptions}, the two within-group cases have the same results. Therefore, the expectation of $\mathcal{D}$ can be expressed as
\begin{equation}\label{eqnEGij-general}
  \begin{gathered}
    E \left( \mathcal{D}_{ij} \right)=
    \begin{cases}
      0, & i=j. \\
      \mu_{\text{within}}, & i,j \in A \text{ or } i,j \in B,\; i\ne j.\\
      \mu_{\text{between}} := \mu_{\text{within}} + \delta_{\mu}, & i\in A,j\in B,\; \text{or } i\in B, j\in A.
    \end{cases}
  \end{gathered}
\end{equation}
with effect size $\delta_{\mu} = 0$ under $H_{0}$ and $\delta_{\mu} > 0$ under $H_{1}$.

Likewise, the covariance $\mathrm{cov}(\mathcal{D}_{ij}, \mathcal{D}_{i'j'})$ can take at most four distinct nonzero values ($\sigma_{1}^{2}$, $\sigma_{2}^{2}$, $\sigma_{3}^{2}$, and $\sigma_{4}^{2}$), depending on the group memberships of $i,j,i',j'$. See Supplementary Text, Section~\ref{apdx-sec:repr-mean-cov} for detailed classifications.

For most practical applications, both $n_{A}$ and $n_{B}$ are large, and $\delta_{\mu}$ is small. Besides, we know that $\sigma_{3}^{2} -\sigma_{1}^{2}=0$ and $\delta_{\sigma^{2}} := \sigma_{4}^{2} -\sigma_{2}^{2} = 0$ when $\delta_{\mu}=0$ (no group difference), therefore it is reasonable to assume that $\delta_{\mu}$ decreases as a function of $n$ and $\delta_{\sigma^{2}} \to 0$ when $n\to \infty$. Based on these considerations, we assume that there exist $\lambda \in (0,1)$, $\uca, \ucb \in \R^{+}$, such that
\begin{equation}
  \label{eq:large-sample}
  \begin{gathered}
    n_{A}/n \to \lambda, \qquad \lim_{n\to \infty} \sqrt{n} \delta_{\mu}(n) = \uca, \quad \lim_{n\to \infty} \sqrt{n} \delta_{\sigma^{2}}(n) = \ucb. \\
    \lim_{n\to \infty} \left( \sigma_{3}^{2}(n) -\sigma_{1}^{2}(n) \right) = 0.
  \end{gathered}
\end{equation}

\begin{theorem}[Universal asymptotic properties for moments]
  \label{thm:main}
  For every distance function and underlying distribution of the networks that satisfy assumptions listed in Section~\ref{sec:notations-assumptions} and Equation~\eqref{eq:large-sample}, the $F$-test defined in Equation~\eqref{eqnTest} has the following asymptotic properties of moments
  \begin{equation}
    \label{eqnasymp-F}
    E\left(F(\mathcal{D}_{n}) \right) =
    \begin{cases}
      1, & \text{under $H_{0}$} \\
      1 + \dfrac{2c_{\mu}\lambda(1-\lambda) \sqrt{n}}{\mu_{\text{within}}}, & \text{under $H_{1}$}
    \end{cases}
  \end{equation}
  \begin{equation}
    \label{eqnasymp-F2}
    \mathrm{var}\left(F(\mathcal{D}_{n})\right) =
    \begin{cases}
      \dfrac{2\sigma_{1}^{2} -4\sigma_{2}^{2}}{ \mu_{\text{within}}^{2} }, & \text{under $H_{0}$}, \\
      \dfrac{4\lambda(1-\lambda) c_{\sigma^2} \sqrt{n}}{\mu_{\text{within}}^{2}}, & \text{under $H_{1}$}.
    \end{cases}
  \end{equation}

  Furthermore, the asymptotic mean and variance of the $F$-test computed from the permuted samples, under either $H_{0}$ or $H_{1}$, matches that of $F(\mathcal{D}_{n})$ under $H_{0}$. That is to say
  \begin{equation}
    \label{eq:asymp-Fperm}
    \begin{split}
      E\left(F(\pi(\mathcal{D}_{n}))\right) = 1, \qquad \mathrm{var}\left(F(\pi(\mathcal{D}_{n}))\right) = \dfrac{2\sigma_{1}^{2} -4\sigma_{2}^{2}}{ \mu_{\text{within}}^{2} }.
    \end{split}
  \end{equation}
\end{theorem}

The proof of Theorem~\ref{thm:main} is highly technical, and we present it in Supplementary Text, Sections~\ref{apdx-sec:large-sample-approx} -- \ref{apdx-sec:exp-var-Fperm}. As a useful by-product of these derivations, we carefully studied the asymptotic mean and covariance matrix of $\mathcal{D}$, which may be useful for other research projects that need statistical properties of the distance matrix.  Graphical illustrations of $E\mathcal{D}$ and $\mathrm{cov}\left( \mathcal{D} \right)$ are provided in Figure S1.

We know that when permutation-based null distributions of $F$-statistics converges to the theoretical null distribution, the permutation test attains 100\% asymptotic relative efficiency (ARE) compared with the corresponding parametric test based on the \emph{oracle null distribution}. In this sense, Theorem~\ref{thm:main} seems insufficient. Unfortunately, we must point out that the limiting distribution of properly standardized $F(\mathcal{D}_{n})$ may not be \emph{normal}, and there is no \emph{universal} limit distribution that works for all network distributions and $d(\cdot, \cdot)$. In this sense, the universal asymptotic properties of the first and second order moments we provide in Equations~\eqref{eqnasymp-F} and \eqref{eqnasymp-F2} are the best possible theoretical results for distance-based permutation $F$-test. This claim will be made evident through an example in the following section.

\subsection{An Example with the  Euclidean Distance}\label{CasePropSect}

Let $\mathcal{D}_{ij} := d^{2}(X_{i}, X_{j})$ be the squared Euclidean distance between two $v$-dimensional random vectors $X_{i}$ and $X_{j}$ with the following multivariate normal distribution
\begin{equation}
  \label{eqnmultvar-norm}
  X_{i} = Z_{i} + \mathbf{m}_{i}, \quad Z_{i} \sim N(0_{v}, I_{v}), \quad \mathbf{m}_{i} =
  \begin{cases}
    0_{v}, & i\in A, \\
    \mathbf{v}, & i\in B.
  \end{cases}
\end{equation}

We are interested in testing $H_{0}: \mathbf{v} = 0_{v}$ against $H_{1}: \mathbf{v} \ne 0_{v}$. With some work (see Section~\ref{apdx-sec:squar-eucl-dist} in Supplementary Text), we are able to show that when $n\to \infty$, the $F$-statistic computed from the original and permuted samples converges to the following distributions
\begin{equation}
  \label{eqnF-asymp-Euclidean}
  F \stackrel{d}{\longrightarrow} \frac{1}{v} \cdot \chi^{2}_{v}\left(\frac{n_{A}n_{B} \delta_{\mu}}{n} \right) = \frac{1}{v}\cdot \chi^{2}_{v}\left(c_{\mu}\lambda(1-\lambda)\sqrt{n} \right).
\end{equation}
\begin{equation}
  \label{eqnFPerm-asymp-Euclidean}
  F(\pi(\mathcal{D}_{n})) \stackrel{d}{\longrightarrow} \frac{1}{v}\cdot \chi^{2}_{v}.
\end{equation}

We see that when the exact form of the distance is given (\textit{e.g.} the Euclidean distance in $\R^{v}$), we are able to derive the asymptotic distribution of $F(\mathcal{D}_{n})$, not just its mean and variance. We are also able to prove that the theoretical and permutation-based null distributions of $F$-statistics are asymptotically equivalent, thus the permutation $F$-test attains $100\%$ ARE relative to a parametric test based on oracle null distribution. However, Equation~\eqref{eqnF-asymp-Euclidean} also shows that, even for the Euclidean distance, the asymptotic distribution of $F(\mathcal{D}_{n})$ is not normal, and more importantly, it depends on $v$, a parameter of the underlying distribution of $X$ (the analogy of networks).  Therefore it is \textbf{impossible} to derive an universal asymptotic distribution of $F(\mathcal{D}_{n})$ that works for all cases.

\section{Simulation}\label{SimuSect}
In this section, the performance of the proposed method and several selected alternatives will be assessed via comprehensive simulation studies. Note that the computational cost of the proposed test is as expensive as $O(Kv^2n^2)$, where $K$ is the number of permutation, $v$ is the vertex size, and $n$ is the total sample size. Thus in the simulation experiments below, we only consider networks with less than 100 vertices and often restrict it to 20. We recognize that in the field of neuroscience, brain networks could become high dimensional (e.g., hundreds of vertices) such that further research work is needed to improve the computing efficiency of our method in the future.

\subsection{Experiment Design}\label{DesignSect}

To evaluate the performance of the proposed method on comparing network topology, in the first scenario, we generate un-weighted graphs from three graph families: Erdos-Renyi model (ER; \cite{erdHos1960evolution}), random bipartite graph (BP), and Barabasi-Albert model (BA; \cite{albert2002statistical}). The ER model is a classical random graph generation tool so it can serve as an ideal control. We choose BP and BA models because both of them can generate graphs that mimic real world networks \citep{albert2002statistical, guillaume2006bipartite} and the graph features are easy to control for comparison purpose. Several experiments have been conducted for comprehensive performance evaluation in terms of both powers and type I errors in the first scenario. These experiments can be categorized into four types of comparisons: (a) ER vs BP; (b) ER vs BA; (c) ER vs ER and (d) BA vs BA. For scenarios 1.a, 1.b and 1.c, the vertex sizes and sample sizes are fixed to be 20 for both groups. In scenario 1.a, the edge generating probabilities are set to be 0.02, 0.04, ..., and 0.3 for ER networks while the parameters are adjusted accordingly in BP networks to make sure that the edges in both groups have the same generating probabilities in average. Similarly in scenario 1.b, the numbers of edges being added at each step are set to be 1, 2, ..., and 10 for BA networks and the edge generating probabilities of ER networks are adjusted accordingly to assure the underlying edge densities are equal to BA networks. In scenario 1.c, the edge generating probabilities for both ER network groups are chosen from 0.1, 0.3, ..., and 0.9. In scenario 1.d, the numbers of edges being added at each step are set to be 1 for both BA network groups while having a vertex size of 20, 50 or 100 for different experiments.

In the second scenario, we extend the comparisons between unweighted networks to weighted networks. For simplicity, the edge weights are assumed to follow Multivariate Normal distributions. Let $\mu$, $\sigma^2$ and $\Sigma$ be the mean, variance and correlation matrix of the multivariate normal distribution, respectively. Here $\sigma^2$ is a single number while $\mu$ and $\Sigma$ are a $|E| \times 1$ vector and a $|E| \times |E|$ matrix, respectively. In the second scenario, there are mainly 3 categories of comparisons: (a) independent edges with different means; (b) dependent edges with different means; (c) independent edges with different densities. All the experiments in this scenario are conducted with the vertex size and sample size being 20 for both groups under comparison. For scenarios 2.a and 2.b, the networks are assumed to be fully connected. The variance $\sigma^2$ is set to be 1. In group 1, the means of the edges $\mu$ are fixed to be 10 while in group 2, $\mu = 8, 8.05, ..., 12$. The only difference is that the edge correlation structure $\Sigma$ is assumed to be $I$ in scenario 2.a and be toeplitz$(1:\frac{1}{190})$ in scenario 2.b. The number of edges for such full connected networks is 190 and the sample size is 20, thus we have
$$\text{Toeplitz}(1:\frac{1}{190}) = \begin{bmatrix}
 1 & \frac{189}{190} & \cdots & \frac{2}{190} & \frac{1}{190} \\
 \frac{189}{190} & 1 & \cdots & \frac{3}{190} & \frac{2}{190} \\
 \vdots & \vdots & \ddots & \vdots & \vdots \\
 \frac{2}{190} &  \frac{3}{190} & \cdots & 1 & \frac{189}{190} \\
 \frac{1}{190} &  \frac{2}{190} & \cdots & \frac{189}{190} & 1
\end{bmatrix}.$$
For scenario 2.c, networks are generated from ER model and $\mu=4$, $\sigma^2=0.25$, $\Sigma=\mathbf{I}$ for both groups. The edge generating probability in group 1 is 0.5 while in group 2 is chosen from 0, 0.05, ..., 1.

In scenario 3, experiments are also conducted on weighted graphs to further examine the effects of different sample sizes and vertex sizes. All the experiments in this scenario use the same settings of $\mu=4$, $\sigma^2=0.25$, $\Sigma=\mathbf{I}$ and the edge generating probabilities are chosen from 0.1, 0.15, and 0.2 for different experiments. Networks are generated from ER model and BP model for two groups, respectively. In scenario 3.a, similar to previous experiments in scenario 2, both sample size and vertex size remain at 20. In scenario 3.b, the vertex size increases to 100; and in scenario 3.c, the sample size increasing to 100.

For all the simulation experiments above, the p-value in each hypothesis testing is calculated based on 1000 permutations. The proposed method is mainly compared with other four selected two-sample testing approaches: i) a two-sample test for graphs based on minimum spanning tree that developed by \cite{Chen2017}; ii) a two-sample test for networks based on Laplacian matrices developed by \cite{ginestet2017hypothesis}; iii) a Wilcoxon rank-sum test that based on a summary statistics of networks named global efficiency \citep{latora2001efficient}; and iv) classical Binomial test (only applied to un-weighted graphs). The empirical powers and type I errors are estimated via the rates of rejecting null hypothesis based on 1000 independent repetitions at a significance level of 0.05.

In the simulation scenarios above, the distance metric in Eq. (\ref{eqnQIMmetric}) is used with $\kappa = 1$. Also, if Eq. (\ref{eqnQIMmetric}) is denoted as $d_{q}= d_{a} \cdot (1+ \kappa \cdot d_{im})$, we can consider a ``plus version'' of the proposed distance metric as $d_{q}= d_{a}+\kappa \cdot d_{im}$, which is of our interest to explore further for performance evaluation. Therefore, in scenario 4, we consider the same experimental settings as in scenario 1; and for both the ``Plus version'' and the original version of the distance metric, we assess the performances of the proposed method with $\kappa = 0, 0.001, 0.01, 0.1, 1, 10, 100, 1000$.

\subsection{Simulation Results}\label{SimresultSect}

The simulation results in scenario 1 are shown in Table~\ref{tabscenario1}. The results of scenarios 1.a and 1.b indicate that, when the underlying edge densities are the same, Ginestet's test and Binomial test may fail to distinguish unweighted networks from different distributions. None of them achieve a power greater than 0.1 while both Chen's test and our proposed test have a power greater than 0.5 in most of the cases. Experiments in scenario 1.a also demonstrate that as edge density increases, the power of the proposed test and Chen's test increases; moreover, our proposed method has a higher power than that of Chen's test. When the edge density is equal to 0.1, the power of our method reaches 0.923 while the power of Chen's test is 0.617. Different from other tests, the global efficiency test fails in scenario 1.a but turn out to work well in 1.b. This may be due to the fact that global efficiency is a good measure for fully connected networks. In scenario 1.a, networks generated from both ER and BP models with low edge densities are not fully connected; however, in scenario 1.b, BA networks tend to be fully connected. In addition, as shown in Table~\ref{tabscenario1}, both Chen's test and our test can achieve a power greater than 0.9 when comparing BA networks with ER networks even for a very low edge density; however, when comparing BP networks with ER networks, both methods has a power less than 0.5 if the edge density is 0.06. Scenario 1.c and 1.d show that both the proposed test and Chen's test can control the type I error rate while other tests may fail in some cases.

In scenario 2, as shown in Table~\ref{tabscenario2}, when a graph has weighted edges and a high edge density, the proposed method and the global efficiency test perform the best under the experiment settings of scenario 2.a, 2.b and 2.c (the powers are greater than 0.9 in all cases). Chen's test still works well while its power becomes slightly lower. For example, in scenario 2.a, when the underlying mean difference is 0.25, our method and the global efficiency test can achieve a power greater than 0.99 while Chen's test reaches a power of 0.86 or higher. Ginestet's test fails to detect the difference in this case, with a power less than 0.01. When the mean difference increases to 1.5, Ginestet's test starts to increase its power around 0.5. In scenario 2.b, when edges are no longer independent but with a correlation structure as toeplitz$(1:\frac{1}{190})$, the powers of all the selected methods decrease. When the mean difference between two groups is 1.05, the powers of the proposed method and the global efficiency test are around 0.97. In scenario 2.c, after the edge densities decreases to around 0.5, the proposed method still has a power greater than 0.9 when the underlying probability is different by 0.05 in two groups. In this case, Chen's test only achieves a power of 0.4 or lower and the global efficiency test performs better with a power greater than 0.93. Among all the three experiments in scenario 2, the type I errors of the proposed method, Chen's test and the global efficiency test are close to 0.05.

In Table~\ref{tabscenario3} for scenario 3, where both the sample size and the vertex size are fixed at 20 with an edge density 0.1, the proposed method and the Chen's test have a power of 0.93 and 0.56, respectively. Increasing either the sample size or the vertex size from 20 to 100 can improve the power of the proposed test and Chen's test to more than 0.98. For Ginestet's test, the power increases from less than 0.1 to more than 0.99 when the sample size increases from 20 to 100; however, increasing the vertex size from 20 to 100 has little effect on its power. On the contrary, the power of the global efficiency test increases from around 0.05 to more than 0.9 when the vertex size increases from 20 to 100 but remains unchanged when the sample size increases. Note that Table~\ref{tabscenario1} and Table~\ref{tabscenario2} only show the selected representative results, please see Supplementary Section~\ref{apdx-sec:additional-results} for more results.

\begin{figure}
{\includegraphics[width=1\textwidth]{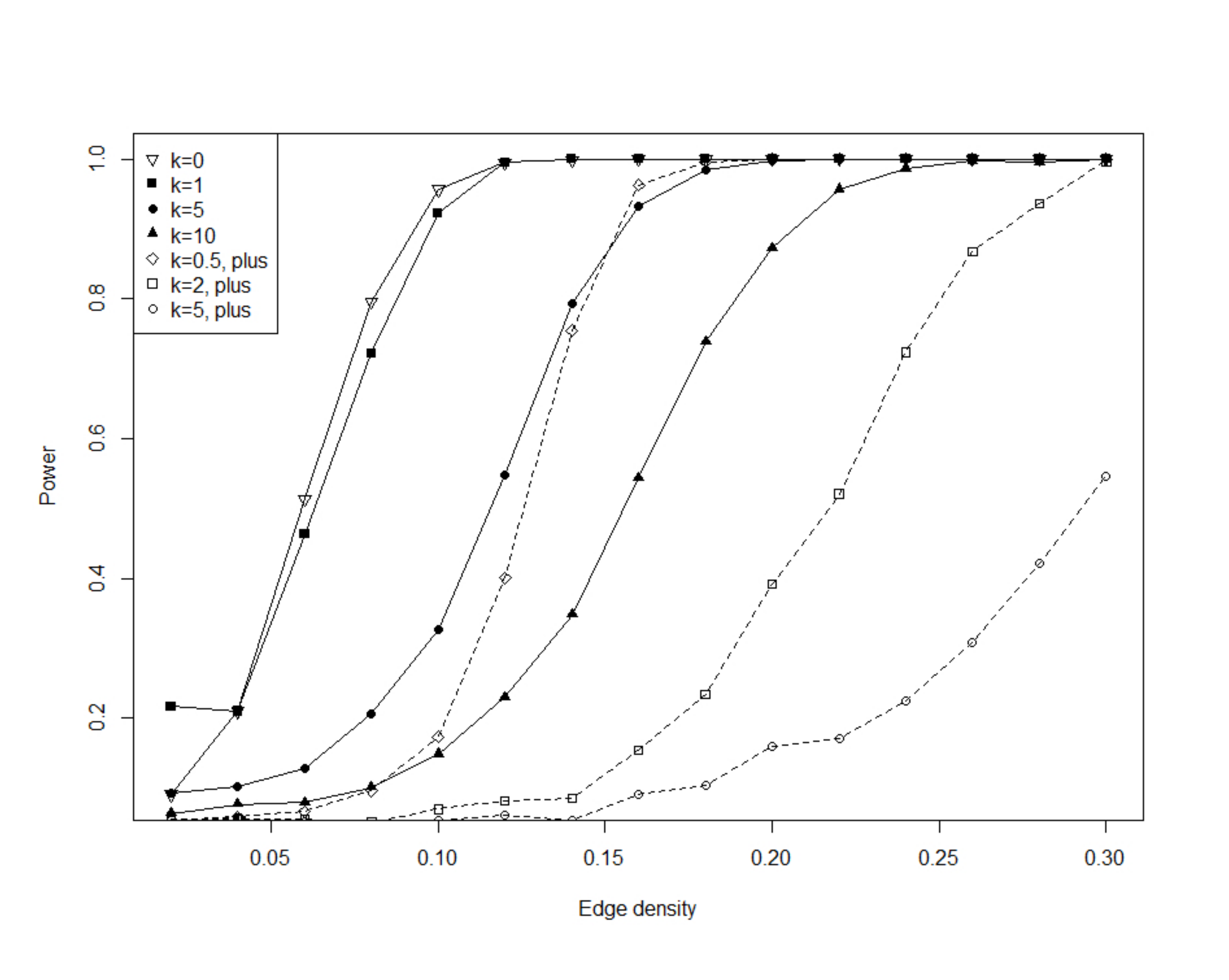}}
\caption{Simulation Results of Scenario 2.
    All of the experiments under this scenario are conducted based on 1000 iterations and calculating the rates of rejecting the null hypothesis under the 0.05 significance level. Two groups of unweighted graphs are generated from two different distributions, ER and BP, with the Sample Size = 20, Vertex Size=20 and same edge densities. Experiment 1.a. examine the power of discriminating ER and BP distributions under different edge densities. }
\label{figSimulationScenario4}
\end{figure}

Simulation results in scenario 4 are shown in Figure~\ref{figSimulationScenario4}. One can tell that a smaller value of the weight parameter $\kappa$ is associated with a greater power for the proposed test. For the proposed method using the original distance metric, when $\kappa = 10$, it achieves a power greater than 0.9 at an edge density around 0.20; while when $\kappa = 5$, an edge density around 0.15 corresponds to a power of 0.9; when $\kappa = 1$, an edge density of 0.10 corresponds to a power of 0.9 or higher for the test. However, using the plus version of the distance metric in the proposed test, when $\kappa = 0.5$ the edge density should be around 0.15 to obtain a power more than 0.9 and when $\kappa$ increases to 2, the minimum edge density for a 0.9 power increases to 0.25. Thus, the proposed distance metric with $(\kappa=1)$ that we used in previous scenarios might be empirically optimal among the different settings we consider.

\section{Real Application}\label{AppSect}

Attention-deficit/hyperactivity disorder (ADHD) is a neurodevelopmental disorder characterized by symptoms of impulsivity, inattention and hyperactivity among children \citep{american2013diagnostic}. It may lead to learning, emotional, social relationship and adaptation problems and thus negatively affect patients' academic achievement, career development, and quality of life. It is estimated that the worldwide prevalence of ADHD is about $5\%$ \cite{polanczyk2007worldwide} and approximately $70\%$ of ADHD children persist into adulthood \cite{lara2009childhood}. Adult ADHD tends to be associated with depression, anxiety disorder, substance abuse, traffic accidents, and crimes \cite{klassen2010adult}. Therefore, ADHD has become an important public health problem \cite{frodl2010comorbidity}. However, the neural substrates associated with ADHD, from both structural and functional perspectives, are not yet well established.

In 2011, the ADHD-200 Consortium organized the ADHD-200 Global Competition on diagnosing individuals with ADHD and made a set of data publicly available for research use. To avoid the discrepancy between different processing pipelines for fMRI data, in this study, a subset of the original ADHD-200 data \citep{milham2012adhd} called "ADHD200-CC200" is used. Available from an open data source, USC Multimodal Connectivity Database (\url{http://umcd.humanconnectomeproject.org}, \cite{brown2012ucla}), the ADHD200-CC200 dataset contains the resting-state fMRIs of 190 ADHD patients and 330 typically developing (TD) controls from five independent neuroimaging scanning sites. The raw fMRI images have been previously processed and converted to resting-state functional connectivity matrices based on 190 brain regions of interest (ROI). Based on the linear correlation between ROIs, each element in a connectivity matrix is a measure of activity dependency between two different brain areas of a patient.

\begin{figure}
{\includegraphics[width=1\textwidth]{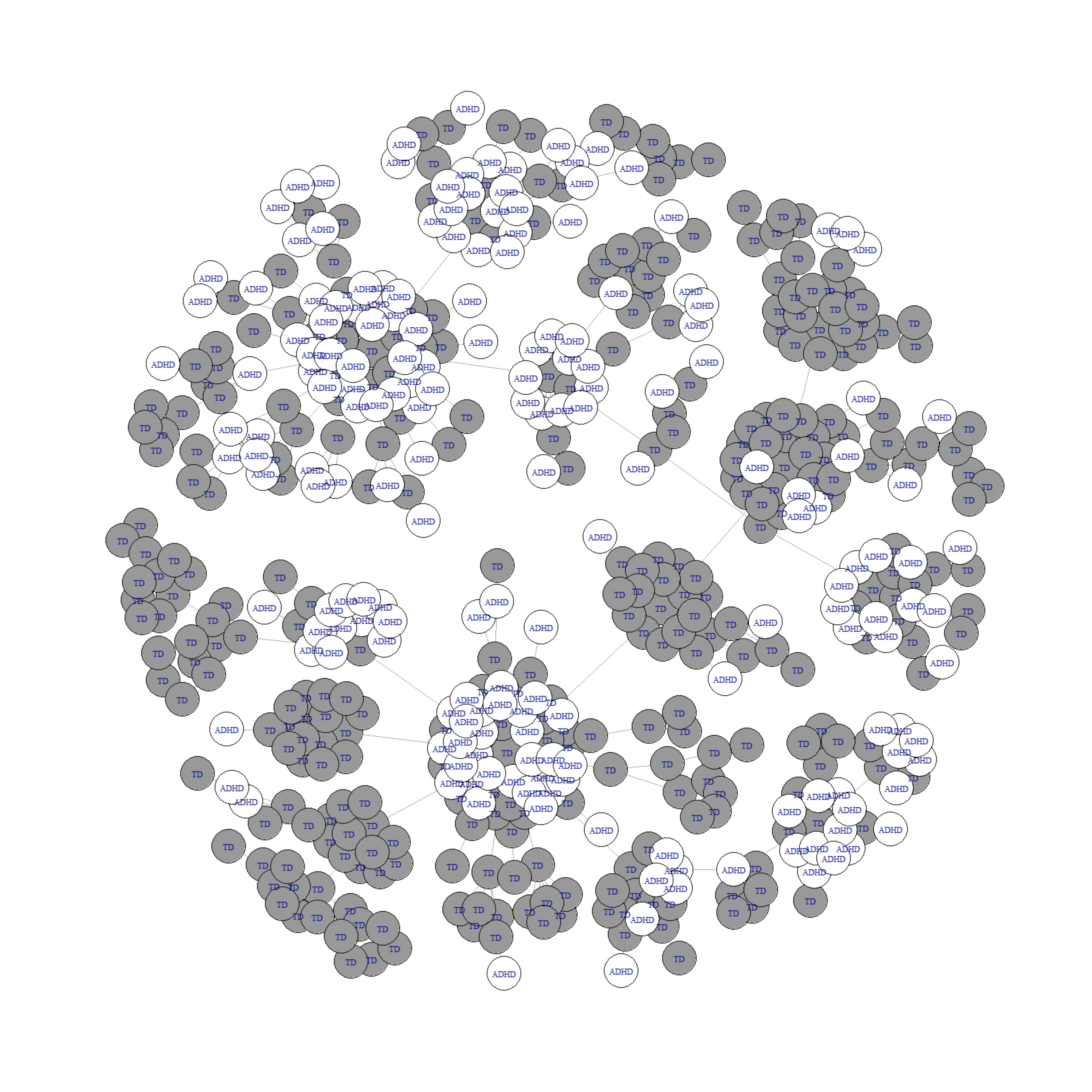}}
\caption{The left figure demonstrates the clustering results of the ADHD200-CC200 dataset by applying MST-KNN directly to the pairwise QIM. The right figure shows whether certain patient has disease or not (Dark green: ADHD group; Yellow: TD group).}
\label{figApplicationADHD200}
\end{figure}

To appreciate the practical usefulness and importance of the proposed method, we visualize the distribution of these network data first. As each subject's data is represented by a connectivity matrix, all the QIM distances can be computed and then used to project a total of 520 brain networks onto a 2-D plane using minimum spanning tree k-nearest neighbor (MST-kNN) \citep{inostroza2008integrated}, as shown in Fig. \ref{figApplicationADHD200}. It is obvious in Fig. \ref{figApplicationADHD200} that ADHD and TD subjects are well mixed together and there is not any notable cluster structure in this figure that can distinguish ADHD from TD subjects. An na\"{\i}ve data science analysis may just stop here and draw the conclusion that the ADHD group is not different from the TD group in terms of brain connectivity network. However, by directly applying our method to the ADHD200-CC200 dataset, a p-value of 0.0055 is obtained when comparing the TD group with the ADHD group. This result suggests that while na\"{\i}ve (but commonly used) machine learning approaches may fail to detect hidden patterns in brain connectivity networks, appropriate statistical (learning) methods can successfully tell the statistical significance of the difference between complex object data distributions.

\section{Conclusion and Discussion}\label{DiscConcSect}
In this study, we propose a new permutation-based $F$-test for comparing two groups of networks, which utilizes both local and global information of networks and has a superior power over other alternatives for network data. Unlike other pioneer work that also addresses this issue, we take the non-Euclidean characteristics of network data into consideration. We also prove that such test always has tight control of type 1 error since it utilizes permutation to conduct hypothesis test. In addition, the asymptotic properties of the mean and variance of the distance matrix and the permutation $F$-statistic under both null and alternative hypotheses have been studied. These results can be used together with analytic tools such as Chebyshev's inequality to obtain certain bounds of statistical power. Somewhat surprisingly, we discovered that no universal asymptotic distribution exist in all situations and demonstrate this by a concrete example. On the other hand, we showed with a concrete example that when the exact form of the distance is given, it is possible to obtain much stronger theoretical conclusions and prove the asymptotic efficiency of the distance-based permutation $F$-test.

We have illustrated in the simulation experiments that the proposed method is less strict about the assumptions and can be applied to wide range of situations. It obtains the highest power or a power as good as those of other alternatives in different scenarios in Section~\ref{SimuSect}, covering networks with weighed or unweighted edges, high or low edge density, independent or correlated edges and so on. Meanwhile, the empirical type I errors can be constantly controlled around 0.05. This may be due to the non-parametric nature of the test formulation, including both the distance metric and the hypothesis testing procedure. As Chen's test is a generic method, it obtains good powers in all different scenarios as expected. But obviously, the proposed method can outperform Chen's method in all the cases presented in Section~\ref{SimuSect}. One possible explanation could be that Chen's test is an Euclidian tool in nature. Other methods, including Ginestet's test, Global efficiency test, and Binomial test,
were found to fail in certain scenarios due to the limitations in their underlying assumptions.

We also recognize that our proposed method might not be applicable to high dimensional cases. One important but less-recognized problem associated with high dimensionality is a phenomenon called {\it hubness} \citep{Radovanovic2010}. That is, even if a data point $\mathbb{A}$ is one of the $k$-nearest neighbors (KNN) of another data point $\mathbb{B}$, it is not necessarily true that $\mathbb{B}$ is also the KNN of $\mathbb{A}$. Such a possible asymmetry in nearest neighbor relations has profound effects on two-sample tests: points in sample $\mathcal{B}$ may find points in sample $\mathcal{A}$ closer than those in $\mathcal{B}$ even if $\mathcal{A}$ and $\mathcal{B}$ are well separate from each other. This problem will occur even if an accurate geodesic distance metric is used. 
It is clear that test statistics which directly employ QIM (e.g.,  $d_q(\hat{\mu}_{\mathcal{A}}, \hat{\mu}_{\mathcal{B}})$) will run into the aforementioned issue; in addition, such test statistics are not shift and scale invariant. To circumvent these difficulties, we can construct test statistics by considering QIM-based mutual remoteness (MR) of data points. As suggested by \cite{Schnitzer2012}, mutual proximity (MP), as one of the global scaling methods, can be used to deal with the hubness problem by considering the fact that two points sharing more nearest neighbors may be closer to each other. For the pooled sample $\mathcal{Z}$, let $H(Z_i)$ ($i=1,\dots,S$) denote the probability distribution of distances from $\mathcal{G}_i^{\mathcal{Z}}$ to all other points in $\mathcal{Z}$; also, let $H(Z_j)$ ($i \neq j$ and $j=1,\dots,S$) denote the distance distribution of another point $\mathcal{G}_j^{\mathcal{Z}}$, and let $H(Z_i, Z_j)$ denote the joint distribution of $Z_i$ and $Z_j$. Note that, although $d_q(\mathcal{G}_i^{\mathcal{Z}}, \mathcal{G}_j^{\mathcal{Z}}) = d_q(\mathcal{G}_j^{\mathcal{Z}}, \mathcal{G}_i^{\mathcal{Z}})$, $H(Z_i < z)$ is likely to be different from $H(Z_j < z)$ because the distance distribution of $\mathcal{G}_i^{\mathcal{Z}}$ may be different from that of $\mathcal{G}_j^{\mathcal{Z}}$. The MP between $\mathcal{G}_i^{\mathcal{Z}}$ and $\mathcal{G}_j^{\mathcal{Z}}$ is defined as follows
  \begin{equation}\label{eqnMPDef}
    MP(\mathcal{G}_i^{\mathcal{Z}}, \mathcal{G}_j^{\mathcal{Z}}) = H \left( Z_i > d_q(\mathcal{G}_i^{\mathcal{Z}}, \mathcal{G}_j^{\mathcal{Z}}) \cap Z_j > d_q(\mathcal{G}_j^{\mathcal{Z}}, \mathcal{G}_i^{\mathcal{Z}}) \right).
  \end{equation}
Without introducing any parametric distribution assumption, the MP can be calculated using the empirical distribution as follows
\begin{equation}\label{eqnMPEmpirical}
 MP(\mathcal{G}_i^{\mathcal{Z}}, \mathcal{G}_j^{\mathcal{Z}}) =  \frac{\left| \left\{ k: d_q(\mathcal{G}_i^{\mathcal{Z}}, \mathcal{G}_k^{\mathcal{Z}}) > d_q(\mathcal{G}_i^{\mathcal{Z}}, \mathcal{G}_j^{\mathcal{Z}}) \right\} \cap \left\{ k: d_q(\mathcal{G}_j^{\mathcal{Z}}, \mathcal{G}_k^{\mathcal{Z}}) > d_q(\mathcal{G}_j^{\mathcal{Z}}, \mathcal{G}_i^{\mathcal{Z}}) \right\} \right|}{S}.
\end{equation}
The definition of MP suggests that the more points are closer neighbors of either  $\mathcal{G}_i^{\mathcal{Z}}$ than $\mathcal{G}_j^{\mathcal{Z}}$ or $\mathcal{G}_j^{\mathcal{Z}}$ than  $\mathcal{G}_i^{\mathcal{Z}}$, the larger the value of MP is. Therefore, our MR is defined as
  \begin{equation}\label{eqnMRDef}
    MR(\mathcal{G}_i^{\mathcal{Z}}, \mathcal{G}_j^{\mathcal{Z}}) =1- MP(\mathcal{G}_i^{\mathcal{Z}}, \mathcal{G}_j^{\mathcal{Z}})
  \end{equation}
to assure that a larger value of MR corresponds to a longer distance between  $\mathcal{G}_i^{\mathcal{Z}}$ and $\mathcal{G}_j^{\mathcal{Z}}$. Conceptually,
replacing $d_q(\mathcal{G}_i^{\mathcal{Z}}, \mathcal{G}_j^{\mathcal{Z}})$ in Eq.~(\ref{eqnTest}) with $MR(\mathcal{G}_i^{\mathcal{Z}}, \mathcal{G}_j^{\mathcal{Z}})$, we can extend our proposed method to high dimension cases. For illustration purpose, a few simulation studies have been performed and the results can be found in Supplementary  Table~\ref{apdx-tab:scenariohigh}, which shows that the high-dimensional extension of our method is promising in a high dimensional setting.

Finally, it should be stressed that, while our proposed framework is established for independent samples of network data only, it provides a solid basis for future research work in novel statistical methodology development for non-Euclidean object data.

\end{doublespace}
\renewcommand{\baselinestretch}{1.4}\normalsize
\bibliographystyle{ECA_jasa}
\bibliography{networkref}

\clearpage\pagebreak\newpage\thispagestyle{empty}


\begin{table}[h!]
\small
  \begin{center}
    \caption{Simulation Results of Scenario 1.
    All of the experiments under this scenario are conducted based on 1000 iterations and calculating the rates of rejecting the null hypothesis under the 0.05 significance level. Two groups of unweighted graphs are generated from two distributions with Sample Size = 20, Vertex Size=20 and same edge densities. Scenario 1.a. examine the power of discriminating ER and BP distributions under different edge densities. Experiment 1.b. shares the same settings with 1.a. except for that the distributions are changed to be ER and BA. Scenario 1.c. examine the type I error of ER's under different edge densities. Scenario 1.d. is similar to 1.c. but using BA instead of ER. Note that in 1.d. once the number of vertices is fixed, the edge density is fixed.}
    \label{tabscenario1}
    \begin{tabular}{|p{2cm}|p{2cm}|p{2cm}|p{2cm}|p{2cm}|p{2cm}|}
      \hline\hline
      \multicolumn{6}{|c|}{Scenario 1.a. Power for unweighted graph, ER vs BP}\\
      \hline
      Edge Density & Proposed Method & Chen Test & Ginestet Test & Global efficiency Test & Binomial Test \\ \hline
      0.06 & 0.463 & 0.255 & 0.051 & 0.029    & 0.048 \\ \hline
      0.10 & 0.923 & 0.617 & 0.060 & $<$0.001 & 0.061 \\ \hline
      0.16 & 1.000 & 0.962 & 0.051 & $<$0.001 & 0.051\\ \hline\hline
      \multicolumn{6}{|c|}{Scenario 1.b. Power for unweighted graph, ER vs BA}\\
      \hline
      Edge Density & Proposed Method & Chen Test & Ginestet Test & Global efficiency Test & Binomial Test \\ \hline
      0.05 & 1.000 & 1.000 & $<$0.001 & 1.000 & 0.005 \\ \hline
      0.35 & 1.000 & 1.000 & 0.010    & 1.000 & 0.003 \\ \hline
      0.38 & 1.000 & 1.000 & 0.028    & 1.000 & 0.006 \\ \hline\hline
      \multicolumn{6}{|c|}{Scenario 1.c. type I error for unweighted graph, ER vs ER}\\
      \hline
      Edge Density & Proposed Method & Chen Test & Ginestet Test & Global efficiency Test & Binomial Test \\ \hline
      0.1 &	0.056 &	0.049 &	$<$0.001 & 0.053 & 0.058 \\ \hline
      0.5 &	0.044 &	0.060 &	$<$0.001 & 0.045 & 0.051 \\ \hline
      0.9 &	0.052 &	0.057 &	$<$0.001 & 0.047 & 0.046 \\ \hline\hline
      \multicolumn{6}{|c|}{Scenario 1.d. type I error for unweighted graph, BA vs BA}\\
      \hline
      Vertex Size & Proposed Method & Chen Test & Ginestet Test & Global efficiency Test & Binomial Test \\ \hline
      20  &	0.055 &	0.055 &	$<$0.001 &	0.042 &	$<$0.001\\ \hline
      50  &	0.041 &	0.051 &	$<$0.001 &	0.036 &	$<$0.001\\ \hline
      100 &	0.055 &	0.049 &	$<$0.001 &	0.046 &	$<$0.001\\ \hline
    \end{tabular}
  \end{center}
\end{table}

\begin{table}[h!]
  \begin{center}
    \caption{Simulation Results of Scenario 2. All of the experiments under this scenario are conducted based on 1000 iterations and calculating the rates of rejecting the null hypothesis under the 0.05 significance level. Two groups of graphs are generated from same distributions, ER, with Sample Size = 20 and Vertex Size=20. Scenario 2.a. examine both the power and type I error under the settings of $\sigma^2$=1, $\Sigma=I$ with full connection (edge density = 1). For one of the group, the average edge weight is fixed as $\mu$=10 while for the other group the group mean edge weights vary from 8 to 12. Scenario 2.b. shares the same settings with 2.a. except for the covariance matrix $\Sigma$ is set to be toeplitz(1:$\frac{1}{190}$) to introduce correlation betwen edge weights. Scenario 2.c. fixes $\mu$=4, $\sigma^2$=0.25, $\Sigma=I$ for both groups while making changes in the edge densities. One group has its edge density equals to 0.5. The other group has its edge densities vary from 0.2 to 0.8.}
    \label{tabscenario2}
    \begin{tabular}{|p{2.5cm}|p{2.5cm}|p{2cm}|p{2cm}|p{2cm}|p{2cm}|}
      \hline\hline
      \multicolumn{6}{|c|}{Scenario 2.a. Edge density=1; $\Sigma=I$}\\
      \hline
      Mean Group 1 (Edge) &	Mean Group 2 (Edge) &	Proposed Method	& Chen Test & Ginestet Test &	Global efficiency Test \\ \hline
      10 &	8.5	  & 1.000 &	1.000 &	0.499	 & 1.000\\ \hline
      10 &	9.75  & 1.000 &	0.862 &	$<$0.001 & 1.000\\ \hline
      10 &	10	  & 0.061 &	0.051 &	$<$0.001 & 0.055\\ \hline
      10 &	10.25 &	1.000 & 0.871 &	$<$0.001 & 1.000\\ \hline
      10 &	11.5  &	1.000 &	1.000 &	0.487	 & 1.000\\ \hline\hline
      \multicolumn{6}{|c|}{Scenario 2.b. Edge density=1; $\Sigma$=toeplitz(1:$\frac{1}{190}$)}\\\hline
      Mean Group 1 (Edge) &	Mean Group 2 (Edge) &	Proposed Method	& Chen Test & Ginestet Test &	Global efficiency Test \\ \hline
      10 &	8.8   &	0.990 &	0.905 &	0.006 &	0.991\\ \hline
      10 &	8.95  &	0.973 &	0.793 &	0.002 &	0.967\\ \hline
      10 &	10	  & 0.039 &	0.055 &	$<$0.001 & 0.039\\ \hline
      10 &	11.05 &	0.976 &	0.817 &	0.007 &	0.972\\ \hline
      10 &	11.2  &	0.992 &	0.899 &	0.010 &	0.992\\ \hline\hline
      \multicolumn{6}{|c|}{Scenario 2.c. Edge density=0.5; $\Sigma=I$}\\\hline
      Edge Density Group 1 & Edge Density Group 2  &	Proposed Method	& Chen Test & Ginestet Test &	Global efficiency Test \\ \hline
      0.5 &	0.4  &	1.000 &	0.998 &	$<$0.001 &	1.000\\ \hline
      0.5 &	0.45 &	0.904 &	0.426 &	$<$0.001 & 	0.938\\ \hline
      0.5 &	0.5	 &  0.054 &	0.054 &	$<$0.001 & 	0.043\\ \hline
      0.5 &	0.55 &	0.910 &	0.126 &	$<$0.001 & 	0.938\\ \hline
      0.5 &	0.6	 &  1.000 &	0.734 &	$<$0.001 & 	1.000\\ \hline
    \end{tabular}
  \end{center}
\end{table}

\begin{table}[h!]
  \begin{center}
    \caption{Simulation Results of Scenario 3. All of the experiments under this scenario are conducted based on 1000 iterations and calculating the rates of rejecting the null hypothesis under the 0.05 significance level. Scenario 3.a. examine the power of each test under the settings of $\mu$=4, $\sigma^2$=0.25, $\Sigma=I$, Sample Size = 20, Vertex Size=20. Two groups of graphs are generated from two different distributions, BP and ER, with the edge density increasing from 0.1 to 0.2. Scenario 3.b. and Scenario 3.c. share the same settings with Scenario 3.a. except for that Scenario 3.b. use 100 as its vertex size and Scenario 3.c. use 100 as its sample size.}
    \label{tabscenario3}
    \begin{tabular}{|p{2.5cm}|p{2cm}|p{2cm}|p{2cm}|p{2cm}|}
      \hline\hline
      \multicolumn{5}{|c|}{Scenario 3.a. Sample Size = 20; Vertex Size=20}\\ \hline
      Edge Density & Proposed method & Chen Test & Ginestet Test & Global efficiency Test \\ \hline
      0.1  & 0.939 & 0.563 & $<$0.001 & 0.06 \\ \hline
      0.15 & 1.000 & 0.977 & $<$0.001 & 0.075 \\ \hline
      0.2  & 1.000 & 0.996 & $<$0.001 & 0.058 \\ \hline\hline
      \multicolumn{5}{|c|}{Scenario 3.b. Sample Size = 20; Vertex Size=100}\\\hline
      Edge Density & Proposed method & Chen Test & Ginestet Test & Global efficiency Test  \\ \hline
      0.1  & 1.000 & 1.000 & $<$0.001 &	0.230 \\ \hline
      0.15 & 1.000 & 1.000 & $<$0.001 &	0.910\\ \hline
      0.2  & 1.000 & 1.000 & $<$0.001 &	1.000 \\ \hline\hline
      \multicolumn{5}{|c|}{Scenario 3.c. Sample Size = 100; Vertex Size=20}\\\hline
      Edge Density & Proposed method & Chen Test & Ginestet Test & Global efficiency Test \\ \hline
      0.1  & 1.000 & 0.980 & 1.000 & 0.080 \\ \hline
      0.15 & 1.000 & 1.000 & 1.000 & 0.050 \\ \hline
      0.2  & 1.000 & 1.000 & 1.000 & 0.050 \\ \hline
    \end{tabular}
  \end{center}
\end{table}


\clearpage\pagebreak\newpage\thispagestyle{empty}

\end{document}